\documentclass[graybox,envcountchap,sectrefs]{svmult}

\usepackage{mathptmx}
\usepackage{helvet}
\usepackage{courier}
\usepackage{braket} 
\usepackage{algorithmicx}
\usepackage{algpseudocode} 
\usepackage{amsfonts}

\usepackage{type1cm}
\usepackage{exercise}
\usepackage{makeidx}         
\usepackage{graphicx}        
\usepackage{multicol}        
\usepackage[bottom]{footmisc}
\usepackage[usenames,dvipsnames,x11names]{xcolor}
 \usepackage{listings}
 \usepackage{epic}
 \usepackage{eepic}
 \usepackage{a4wide}
 \usepackage{color}
 \usepackage{amsmath}
 \usepackage{amssymb}
 \usepackage[T1]{fontenc}
 \usepackage{cite} 
 \usepackage{shadow}
 \usepackage{hyperref}
 \usepackage{bezier}
 \usepackage{pstricks}
\usepackage{textcomp,type1ec,pdfpages}
\usepackage{bera}

\makeindex             

\usepackage{hyperref}

\begin{document}

\title{Integrating a Computational Perspective in Physics Courses}
\author{Marcos Daniel Caballero and Morten Hjorth-Jensen}
\institute{Marcos Daniel Caballero \at Department of Physics and Astronomy, Michigan State University, East Lansing, 48824 Michigan, USA and
Department of Physics and Center for Computing in Science Education, University of Oslo, N-0316 Oslo, Norway, \email{caballero@pa.msu.edu} \and
Morten Hjorth-Jensen \at Department of Physics and Astronomy and National Superconducting Cyclotron Laboratory, Michigan State University, East Lansing, 48824 Michigan, USA and Department of Physics and Center for Computing in Science Education, University of Oslo, N-0316 Oslo, Norway, \email{hjensen@frib.msu.edu}}

\maketitle

\abstract{In this contribution we discuss how to develop a physics curriculum for undergraduate students that includes computing as a central element. Our contribution starts with a definition of computing and pertinent learning outcomes and assessment studies and programs. We end with a discussion on how to implement computing in various physics courses by presenting our experiences from Michigan State University in the USA and the University of Oslo in Norway.}

\section{Introduction}

Many important recent advances in our understanding of the physical
world have been driven by large-scale computational modeling and data
analysis, for example, the 2012 discovery of the Higgs boson, the 2013
Nobel Prize in chemistry for computational modeling of molecules, and
the 2016 discovery of gravitational waves.  Given the ubiquitous use
in science and its critical importance to the future of science and
engineering, scientific computing  plays a central role in scientific
investigations and is critical to innovation in most domains of our
lives. It underpins the majority of today's technological, economic
and societal feats. We have entered an era in which huge amounts of
data offer enormous opportunities. \href{{http://pathways.acm.org/executive-summary.html}}{By 2020, it is also expected that one out of every two
jobs in the STEM (Science, Technology, Engineering and Mathematics) fields will be in
computing}
(Association for Computing Machinery, 2013, \cite{ACM2013}).

These developments, needs and future challenges, as well as the
developments that are now taking place within quantum computing,
quantum information theory and data driven discoveries (data analysis and
machine learning) will play an essential role in shaping future
technological developments. Most of these developments require true
cross-disciplinary approaches and bridge a vast range of
temporal and spatial scales and include a wide variety of physical
processes. To develop computational tools for such complex
systems that give physically meaningful insights requires a deep
understanding of approximation theory, high performance computing, and
domain specific knowledge of the area one is modeling.

Computing competence represents a central element in scientific
problem solving, from basic education and research to essentially
almost all advanced problems in modern societies. These
competencies are not limited to STEM fields only. The statistical
analysis of big data sets and how to use machine learning algorithms
belong to the set of tools needed by almost all disciplines,
spanning from the Social Sciences, Law, Education to the traditional
STEM fields and Life Science.  Unfortunately, many of our students at
both the undergraduate and the graduate levels are unprepared to use
computational modeling, data science, and high performance computing,
skills that are much valued by a broad range of employers. This lack of preparation is most certainly no fault of our students, but rather a broader issue associated with how departments, colleges, and universities are keeping up with the demands of these high-tech employers. It is through this integrated computational perspective that we aim to address this.
Furthermore, although many universities do offer compulsory
programming courses in scientific computing, and physics departments
offer one or more elective courses in computational physics, there is
often not a uniform and coherent approach to the development of
computing  competencies and computational thinking. This has
consequences for a systematic introduction and realization of
computing  skills and competencies and pertaining learning outcomes.

The aim of this contribution is to present examples on how to
introduce a computational perspective in basic undergraduate physics
courses, basing ourselves on experiences made at the University of
Oslo in Norway and now also at Michigan State University in the
USA. In particular, we will present the \textbf{Computing in Science
Education} project from the University of Oslo \cite{CSEUiO}, a project which has
evolved into a Center of Excellence in Education, the \href{{http://www.mn.uio.no/ccse/english/}}{Center for
Computing in Science
Education} \cite{CCSEUiO}. Similar initiatives
and ideas are also being pursued at Michigan State University.  The
overarching aim is to strengthen the computing  competencies of
students, with key activities such as the establishment of learning
outcomes, how to develop assessment programs and course
transformations by including computational projects and exercises in a
coherent way. The hope is that these initiatives can also lead to a
better understanding of the scientific method and scientific reasoning
as well as providing new and deeper insights about the underlying physics that governs a system.

This contribution is organized as follows. After these introductory
remarks, we present briefly in the next section what we mean by computing and present
possible learning outcomes that could be applied to a bachelor's degree
program in physics (Sec. \ref{sec:competencies}), which are distinguished as more general competencies and course-specific ones.  In Sec. \ref{sec:learingoutcomes}, we discuss possible paths on how to include and implement
computational elements in central undergraduate physics courses. We discuss briefly
how to assess various learning outcomes and how to develop a research program around this.
Several examples that illustrate the links between the learning outcomes and specific mathematics and physics courses
are discussed in Sec. \ref{sec:examples}.
Finally, in the last section we present our conclusions and perspectives.

\section{Computing competencies} \label{sec:competencies}

The focus of this article is on computing competencies and how
these help in enlarging the body of tools available to students and
scientists alike, going well beyond classical tools taught in standard
undergraduate courses in physics and mathematics. We will claim through various
examples that computing  allows for a more generic handling of
problems, where focusing on algorithmic aspects results in deeper
insights about scientific problems.

With \textbf{Computing } we will mean solving scientific problems
using all possible tools, including symbolic computing, computers and
numerical algorithms, experiments (often of a numerical character) and
analytical paper and pencil solutions. We will thus, deliberately,
avoid a discussion of computing and computational physics in
particular as something separate from theoretical physics and
experimental physics.  It is common in the scientific literature to
encounter statements like \emph{Computational physics now represents
  the third leg of research alongside analytical theory and
  experiments}. In selected contexts where say high-performance topics
or specific computational methodologies play a central role, it may be
meaningful to separate analytical work from computational studies. We
will however argue strongly, in particular within an educational
context, for a view where computing means solving scientific problems
with all possible tools. Through various examples in this article we
will show that a tight connection between standard analytical work,
combined with various algorithms and a computational approach, can
help in enhancing the students' understanding of the scientific
method, hopefully providing deeper insights about the physics (or
other disciplines). Whether and how we achieve these outcomes is the
purpose of research in computational physics education.

The power of the scientific method lies in identifying a given problem
as a special case of an abstract class of problems, identifying
general solution methods for this class of problems, and applying a
general method to the specific problem (applying means, in the case of
computing, calculations by pen and paper, symbolic computing, or
numerical computing  by ready-made and/or self-written software).

This generic view on problems and methods is particularly important for
understanding how to apply available generic software to solve a
particular problem.  Algorithms involving pen and paper are
traditionally aimed at what we often refer to as continuous models, of
which only few can be solved analytically. The number of important
differential equations in physics that can be solved analytically are
rather few, limiting thereby the set of problems that can be addressed
in order to deepen a student's insights about a particular physics
case. On the other hand, the application of computers calls for
approximate discrete models.  Much of the development of methods for
continuous models are now being replaced by methods for discrete
models in science and industry, simply because we can address much
larger classes of problems with discrete models, often also by simpler
and more generic methodologies.  In Sec. \ref{sec:examples} we will present
several examples thereof. A typical case is that where an eigenvalue
problem can allow students to study the analytical solution as well as
moving to an interacting quantum mechanical case where no analytical
solution exists. By merely changing the diagonal matrix elements, one
can solve problems that span from classical mechanics and fluid
dynamics to quantum mechanics and statistical physics.  Using
essentially the same algorithm one can study physics cases that
are covered by several courses, allowing teachers to focus
more on the physical systems of interest.

There are several advantages in  introducing computing in basic physics
courses. It allows physics teachers to bring important elements of
scientific methods at a much earlier stage in our students'
education. Many advanced simulations used in physics research can
easily be introduced, via various simplifications, in introductory
physics courses, enhancing thereby the set of problems studied by the
students (see Sec. \ref{sec:examples}). Computing gives university
teachers a unique opportunity to enhance students' insights about
physics and how to solve scientific problems. It gives the
students the skills and abilities that are asked for by
society. Computing allows for solving more realistic problems earlier
and can provide an excellent training of creativity as well as enhancing the
understanding of abstractions and generalizations. Furthermore,
computing can decrease the need for special tricks and tedious
algebra, and shifts the focus to problem definition, visualization,
and "what if" discussions. Finally, if the setup of undergraduate
courses is properly designed, with a synchronization with mathematics
and computational science courses, computing can trigger further
insights in mathematics and other disciplines.

\section{Learning Outcomes and Assessment Programs}\label{sec:learingoutcomes}

An essential element in designing a synchronization of computing in
various physics (and other disciplines as well) courses is a proper
definition of learning outcomes, as well as the development of
assessment programs and possibly a pertinent research program on physics education.
Having a strong physics education
group that can define a proper research program is an essential part
of such an endeavor. Michigan State University has a strong physics
education group involved in such research programs. Similarly, the
University of Oslo, with its recently established center of excellence
in Education \cite{CCSEUiO}, has started to define  a research program that aims at assessing
the relevance and importance of
computing in science education.

Physics, together with basic mathematics and computational science
courses, is at the undergraduate level presented in a very homogeneous
way worldwide.  Most universities offer more or the less the same
topics and courses, starting with Mechanics and Classical Mechanics,
Waves, Electromagnetism, Quantum physics and Quantum Mechanics and
ending with Statistical physics. Similarly, during the last year of
the Bachelor's degree one finds elective courses on computational
physics and mathematical methods in physics, in addition to a
selection of compulsory introductory laboratory courses. Additionally, most physics undergraduate programs have now a compulsory
introductory course in scientific programming offered by the computer science department. Here, one encounters
frequently Python as the default programming language.  Moreover, one
finds almost the same topics covered by the basic mathematics courses
required for a physics degree, from basic calculus to linear algebra,
differential equations and real analysis. Many mathematics departments
and/or computational science departments offer courses on numerical
mathematics that are based on the first course in programming.

These developments have taken place during the last decade and several
universities are attempting to include a more coherent
computational perspective to our basic education. In order to achieve this, it is important to develop a
strategy where the introduction of computational elements are properly
synchronized between physics, mathematics, and computational science
courses. This would allow physics teachers to focus more on the relevant
physics. The development of learning outcomes plays a central role in this work.  An
additional benefit of properly-developed learning outcomes is the
stimulation of cross-department collaborations as well as an increased
awareness about what is being taught in different courses.  Here we
list several possibilities, starting with some basic algorithms and topics that
can be taught in mathematics and computational science courses. We end
with a discussion of possible learning outcomes for central
undergraduate physics courses

\subsection{General Learning Outcomes for Computing Competence.}

Here we present some high-level learning outcomes that we expect
students to achieve through comprehensive and coordinated instruction
in numerical methods over the course of their undergraduate
program. These learning outcomes are different from specific learning
goals in that the former reference the end state that we aim for
students to achieve. The latter references the specific knowledge,
tools, and practices with which students should engage and discusses
how we expect them to participate in that work.

Numerical algorithms form the basis for solving science and
engineering problems with computers. An understanding of algorithms
does not itself serve as an understanding of computing, but it is a
necessary step along the path. Through comprehensive and coordinated
instruction, we aim for students to have developed a deep understanding of:

\begin{itemize}
\item the most fundamental algorithms for linear algebra, ordinary and partial differential equations, and optimization methods;

\item numerical integration including Trapezoidal and Simpson's rule, as well as multidimensional integrals;

\item random numbers, random walks, probability distributions, Monte Carlo integration and Monte Carlo methods;

\item root finding and interpolation methods;

\item machine learning algorithms; and

\item statistical data analysis and handling of data sets.

\end{itemize}

Furthermore, we aim for students to develop:

\begin{itemize}

\item a working knowledge of advanced algorithms and how they can be accessed in available software;

\item an understanding of approximation errors and how they can present themselves in different problems; and

\item the ability to apply fundamental and advanced algorithms to classical model problems as well as real-world problems as well to assess the uncertainty of their results.
\end{itemize}

Later courses should build on this foundation as much as possible. In designing learning outcomes and course contents, one should make
sure that there is a progression in the use of mathematics, numerical
methods and programming, as well as the contents of various physics
courses.  This means also that teachers in other courses do not need
to use much time on numerical tools since these are naturally included
in other courses.

\subsubsection{Learning Outcomes for Symbolic Computing}
Symbolic computing  is a helpful tool for addressing certain classes of
problems where a functional representation of the solution (or part of
the solution) is needed. Through engaging with symbolic computing
platforms, we aim for students to have developed:

\begin{itemize}
\item a working knowledge of at least one computer algebra system (CAS);

\item the ability to apply a CAS to perform classical mathematics including calculus, linear algebra and differential equations; and

\item the ability to verify the results produced by the CAS using some other means.
\end{itemize}

\subsubsection{Learning Outcomes for Programming}
Programming is a necessary aspect of learning computing  for science
and engineering. The specific languages and/or environments that
students learn are less important than the nature of that learning
(i.e., learning programming for the purposes of solving science
problems). By numerically solving science problems, we expect students
to have developed (these are possible examples):

\begin{itemize}
\item an understanding of programming in a high-level language (e.g., MATLAB, Python, R);

\item an understanding of programming in a compiled language (e.g., Fortran, C, C++);

\item the ability to to implement and apply numerical algorithms in reusable software that acknowledges the generic nature of the mathematical algorithms;

\item a working knowledge of basic software engineering elements including functions, classes, modules/libraries, testing procedures and frameworks, scripting for automated and reproducible experiments, documentation tools, and version control systems (e.g., Git); and

\item an understanding of debugging software, e.g., as part of implementing comprehensive tests.
\end{itemize}

\subsubsection{Learning Outcomes for Mathematical Modeling}
Preparing a problem to be solved numerically is a critical step in making progress towards an eventual solution. By providing opportunities for students to engage in modeling, we aim for them to develop the ability to solve real problems from applied sciences by:

\begin{itemize}
\item deriving computational models from basic principles in physics and articulating the underlying assumptions in those models;

\item constructing models with dimensionless and/or scaled forms to reduce and simplify input data; and

\item interpreting the model's dimensionless and/or scaled parameters to increase their understanding of the model and its predictions.
\end{itemize}

\subsubsection{Learning Outcomes for Verification}
Verifying a model and the resulting outcomes it produces are essential elements to generating confidence in the model itself. Moreover, such verifications provide evidence that the work is reproducible. By engaging in verification practices, we aim for students to develop:

\begin{itemize}
\item an understanding of how to program testing procedures; and

\item the knowledge of testing/verification methods including the use of:
\begin{itemize}

  \item exact solutions of numerical models,

  \item classical analytical solutions including asymptotic solutions,

  \item computed asymptotic approximation errors (i.e., convergence rates), and

  \item unit tests and step-wise construction of tests to aid debugging.
\end{itemize}

\end{itemize}

\subsubsection{Learning Outcomes for Presentation of Results}
The results of a computation need to be communicated in some format (i.e., through figures, posters, talks, and other forms of written and oral communication). Computation affords the experience of presenting original results quite readily. Through their engagement with presentations of their findings, we aim for students to develop:

\begin{itemize}
\item the ability to make use of different visualization techniques for different types of computed data;

\item the ability to present computed results in scientific reports and oral presentations effectively; and

\item a working knowledge of the norms and practices for scientific presentations in various formats (i.e., figures, posters, talks, and written reports).
\end{itemize}

The above learning goals and outcomes are of a more generic character. What follows here are specific
algorithms that occur frequently in scientific problems. The implementation of these algorithms in various physics courses, together with problem and project solving, is a way to implement large fractions of the above learning goals.

\subsection{Central Tools and Programming Languages}
We will strongly recommend that Python is used as the high-level
 programming language. Other high-level environments like Mathematica
 and Matlab can also be presented and offered as special courses. This
 means that students can apply their knowledge from the basic programming course offered by most universities.
Many university courses in programming  make use of Python, and extend their computational knowledge in
 various physics classes. We recommend  that the following
 tools are used:
\begin{enumerate}
\item \href{{http://jupyter.org/}}{jupyter and ipython notebooks};

\item version control software like \href{{https://git-scm.com/}}{git} and repositories like \href{{https://github.com/}}{GitHub} and \href{{https://gitlab.com/}}{GitLab};

\item other typsetting tools like {\LaTeX}; and

\item unit tests and using existing tools for unit tests. \href{{https://docs.python.org/2/library/unittest.html}}{Python has extensive tools for this.}
\end{enumerate}

The notebooks can be used to hand in exercises and projects. They can
provide the students with experience in presenting their work in the
form of scientific/technical reports.

Version control software allows teachers to bring in reproducibility
of science as well as enhancing collaborative efforts among
students. Using version control can also be used to help students
present benchmark results, allowing others to verify their
results. Unit testing is a central element in the development of
numerical projects, from microtests of code fragments, to intermediate
merging of functions to final tests of the correctness of a code.

\subsection{Specific Algorithms for Basic Physics Courses}

For a bachelor's degree in physics, it is now more and more common to require a compulsory
programming course, typically taught during the first two years of
undergraduate studies. The programming course, together with
mathematics courses, lay the foundation for the use of computational
exercises and projects in various physics courses. Based on this
course, and the various mathematics courses included in a physics
degree, there is a unique possibility to incorporate computational
exercises and projects in various physics courses, without taking away
the attention from the basic physics topics to be covered.

What follows below is a suggested list of possible algorithms which could be included in central physics courses. The list is by no means exhaustive and is mainly meant as a
guideline of what can be included. The examples we discuss in Sec. \ref{sec:examples}, illustrate how these algorithms can be included in courses like mechanics, quantum physics/mechanics, statistical and thermal physics and electromagnetism. These are all core courses in  a typical bachelor's degree in physics.

\subsection{Central Algorithms}
\begin{itemize}
\item Ordinary differential equations
\begin{enumerate}

  \item Euler, modified Euler, Verlet and Runge-Kutta methods with applications to problems in courses on electromagnetism, methods for theoretical physics, quantum mechanics and mechanics.

\end{enumerate}

\item Partial differential equations
\begin{enumerate}

  \item Diffusion in one and two dimensions (statistical physics), wave equation in one and two dimensions. These are examples of physics cases which could apper in courses on  mechanics, electromagnetism, quantum mechanics, methods for theoretical physics and Laplace's and Poisson's equations in  a course on electromagnetism.

\end{enumerate}

\item Numerical integration
\begin{enumerate}

  \item Trapezoidal and Simpson's rule and Monte Carlo integration. Here one can envision applications in statistical physics, methods of theoretical physics, electromagnetism and quantum mechanics.

\end{enumerate}

\item Statistical analysis, random numbers, random walks, probability distributions, Monte Carlo integration and Metropolis algorithm. These are algorithms with important applications to statistical physics and laboratory courses.

\item Linear Algebra and eigenvalue problems.
\begin{enumerate}

  \item Gaussian elimination, LU-decomposition, eigenvalue solvers, and iterative methods like  Jacobi or Gauss-Seidel for systems of linear equations. These algorithms are important for several courses, classical mechanics, methods of theoretical physics, electromagnetism and quantum mechanics.

\end{enumerate}

\item Signal processing
\begin{enumerate}

  \item Discrete (fast) Fourier transforms, Lagrange/spline/Fourier interpolation, numeric convolutions {\&} circulant matrices, filtering. Here we can think of applications in electromagnetism, quantum mechanics, and experimental physics (data acquisition)

\end{enumerate}

\item Root finding techniques, used in methods for theoretical physics, quantum mechanics, electromagnetism and mechanics.

\item Machine Learning algorithms and Statistical Data Analysis, relevant for laboratory courses
\end{itemize}

In order to achieve a proper pedagogical introduction of these
algorithms, it is important that students and teachers see how these
algorithms are used to solve a variety of physics problems. The same
algorithm, for example the solution of a second-order differential
equation, can be used to solve the equations for the classical
pendulum in a mechanics course or the (with a suitable change of
variables) equations for a coupled RLC circuit in the electromagnetism
course. Similarly, if students develop a program for studies of
celestial bodies in the mechanics course, many of the elements of such
a program can be reused in a molecular dynamics calculation in a
course on statistical physics and thermal physics. The two-point
boundary value problem for a buckling beam (discretized as an
eigenvalue problem) can be reused in quantum mechanical studies of
interacting electrons in oscillator traps, or just to study a particle
in a box potential with varying depth and extension. We discuss some
selected examples in section \ref{sec:examples}. Our coming texbook
\cite{DannyMortenBook} will contain a more exhaustive discussion of these, combined with
a more detailed list of examples and a proper discussion of
learning outcomes and possible assessment programs.

In order to aid the introduction of computational exercises and
projects, there is a strong need to develop educational resources.
Physics is an old discipline with a large wealth of established analytical exercises and
projects. In fields like mechanics, we have centuries of pedagogical
developments with a strong emphasis on developing analytical
skills. The majority of physics teachers are well familiar with this approach.
In order to see how computing  can enlarge this body of exercises
and projects, and hopefully add additional insights to the physics
behind various phenomena, we find it important to develop a large body
of computational examples.
The
\href{{http://www.compadre.org/picup/}}{PICUP project}, Partnership for
Integration of Computation into Undergraduate physics, develops such
\href{{http://www.compadre.org/PICUP/resources/}}{resources for teachers and students on the integration of
computational material} \cite{PICUP}.
We strongly recommend these resources.

\subsubsection{Advanced Computational Physics Courses}
Towards the end of undergraduate studies it is useful to offer a course which focuses on more advanced algorithms and presents compiled languages like C++ and Fortran, languages our students will meet in actual research.
Furthermore, such a course should offer more advanced projects which train the students in actual research, developing more complicated programs and working on larger projects.

\subsection{Physics Education Research and Computing in Science Education}
The introduction of computational elements in the various courses should be, if possible,  strongly integrated with ongoing research on physics education.
The Physics and Astronomy department at MSU is in a unique position due to its strong research group in physics education, the \href{{http://www.pa.msu.edu/research/physics-education-lab}}{PERL group} \cite{PERLMSU}. Together with the Center for Computing in Science Education at the University of Oslo \cite{CCSEUiO}, we are now in the process
of establishing new assessments
and assessment methods that address several issues associated with
integrating computing into science courses. The issues include but
are not limited to how well students learn computing, what new
insights students gain about the specific science through computing,
and how students' affective states (e.g., motivation to learn,
computational self-efficacy) are affected by computing . Broadly
speaking, these assessments should provide deeper insights into the
integration of computing  in science education in general as well as
provide a structured framework for assessment of our efforts and a
basis for systematic studies of student learning.

The central questions that our research must address are
\begin{enumerate}
\item how can we assess the effect of integrating computing  into science curricula on a variety of learned-centered constructs including computational thinking, motivation, self-efficacy and science identity formation,
\item how should we structure assessments to ensure valid, reliable and impactful assessment, which provides useful information to our program and central partners, and finally
\item how can the use of these structured assessments improve student outcomes in teacher-, peer-, and self-assessment.
\end{enumerate}

Addressing these questions requires a combination of qualitative
techniques to construct the focus of these assessments, to build
assessment items and to develop appropriate assessment methods, and
quantitative techniques, including advanced statistical analysis to
ensure validity and reliability of the proposed methods as well as to
analyze the resulting data.

The learning objectives and learning outcomes for computational
methods developed as part of the first objective form parts of the
basis for the assessment program, and we will also investigate the
assessment of non-content learning goals such as self-efficacy and
identity formation. Identifying and investigating the role of such non-content
factors will be critical to support all students in achieving our computational
learning goals.

The effect of integration of computational methods into basic science
courses have been sparsely studied, primarily because the practice is
sparse. Further progress depends now on the development of assessments
that can be used for investigative, comparative and/or longitudinal
studies and to establish best practices in this emerging field.  Some
assessments will be developed for specific courses, but we will aim
for broad applicability across institutions.

\section{Examples on how to Include computing in Physics Undergraduate Programs}\label{sec:examples}

Having defined possible learning outcomes, we would like now to
present some examples which reflect the discussions above. These
examples are mainly taken from various courses at the University of
Oslo, although some of them have been used at Michigan State University. Since 2003, first via the
\href{{http://www.mn.uio.no/ccse/english/people/index.html}}{Computing
  in Science Education project} \cite{CSEUiO} and now through the recently
established center of excellence in education
\href{{http://www.mn.uio.no/ccse/english/}}{Center for computing in
  Science Education} \cite{CCSEUiO}, computing has been introduced across
disciplines in a synchronized way.

Central elements here are a compulsory programming course with a
strong mathematical flavour. This course gives a solid foundation in
programming as a problem solving technique in mathematics. The line of
thought is that when solving mathematical problems numerically, this should enhance
algorithmic thinking, and thereby the students' understanding of the
scientific process.  Secondly, mathematics is at least as important as
before, but should be supplemented with development, analysis,
implementation, verification and validation of numerical
methods. Finally, these methods are used in modeling and problem
solving with numerical methods and visualisation, as well as
traditional methods in various science courses, from the physical
sciences to life science.

Crucial ingredients for the success of the computing in Science
Education project has been the support from governing bodies as well
as extensive cooperations across departmental boundaries. And finally,
the willingness of several university teachers and researchers to give
priority to teaching reforms and course transformations.

In addition to the above, over the years we have coordinated the use
of computational exercises and numerical tools in most undergraduate
courses. Furthermore, via the computing in Science Education project
and now the Center for computing in Science Education, we help in
updating the scientific staff's competence on computational aspects
and give support (scientific, pedagogical and financial) to those who
wish to revise their courses in a computational direction. This may
include the organization of courses for university teachers. Summer
students aid in developing and introducing computational exercises and
several new textbooks have been developed, from the basic mechanics
course to a course in statistical physics \cite{AMS2015, AIV2018,AMSDS2019}.

\subsection{The Physics Undergraduate Program at the University of Oslo.}
The layout of the physics bachelor's degree program at the University of Oslo is shown in table \ref{tab:FAUiO}.
\begin{table}
\caption{The bachelor's degree program in physics at the University of Oslo, Norway}\label{tab:FAUiO}
\begin{footnotesize}
\begin{tabular}{|l|l|l|l|}
\hline
\multicolumn{1}{|l}{ 6th Semester } & \multicolumn{1}{|l|}{ Elective } & \multicolumn{1}{|l|}{ Elective } & \multicolumn{1}{|l|}{ Elective } \\
\hline
5th Semester & FYS2160 Statistical physics & FYS3110 Quantum Mechanics             & Elective                                                        \\
\hline
4th Semeters & FYS2130 Waves and Motion    & FYS2140 Quantum physics               & FYS2150 physics Laboratory                                      \\
\hline
3rd Semester & FYS1120 Electromagnetism    & MAT1120 Linear Algebra                & AST2000 Intro to Astrophysics                            \\
\hline
2nd Semester & FYS-MEK1100 Mechanics       & MEK1100 Vector Calculus               & MAT1110 Calculus and Linear Algebra                             \\
\hline
1st Semester & MAT 1100 Calculus           & MAT-INF1100 Modeling and Computations & IN1900 Intro to Programming \\
\hline
Credits      & 10 ECTS                     & 10 ECTS                               & 10 ECTS                                                         \\
\hline
\end{tabular}
\end{footnotesize}
\end{table}

In the first semester the students encounter the first level of syncronization between computing courses  and mathematics courses.
As an example, consider the numerical evaluation of an integral by  the trapezoidal rule. Integral calculus is typically discussed first in the calculus course MAT1100.
Thereafter, the algorithm for computing  the  integral using the trapezoidal rule for an interval $x \in [a,b]$
\[
  \int_a^b(f(x) dx \approx \frac{1}{2}\left [f(a)+2f(a+h)+\dots+2f(b-h)+f(b)\right],
\]
is discussed and developed in MAT-INF1100, the modeling and
computations course that serves as an intermediate step between the
standard calculus course and the programming course. Finally, the
algorithm is implemented in IN1900, introduction to programming with
scientific applications.  We show here a typical Python code which
exemplifies this.

\begin{lstlisting}
from math import exp, log, sin
def Trapez(a,b,f,n):
   h = (b-a)/float(n)
   s = 0
   x = a
   for i in range(1,n,1):
       x = x+h
       s = s+ f(x)
   s = 0.5*(f(a)+f(b)) +s
   return h*s

def f1(x):
    return exp(-x*x)*log(1+x*sin(x))

a = 1;  b = 3; n = 1000
result = Trapez(a,b,f1,n)
print(result)
\end{lstlisting}
Here we have defined an integral given by
\[
I=\int_1^3 dx \exp{(-x^2)}\log{(1+x\sin{(x)})}.
\]

\begin{figure}
\includegraphics[scale=0.8]{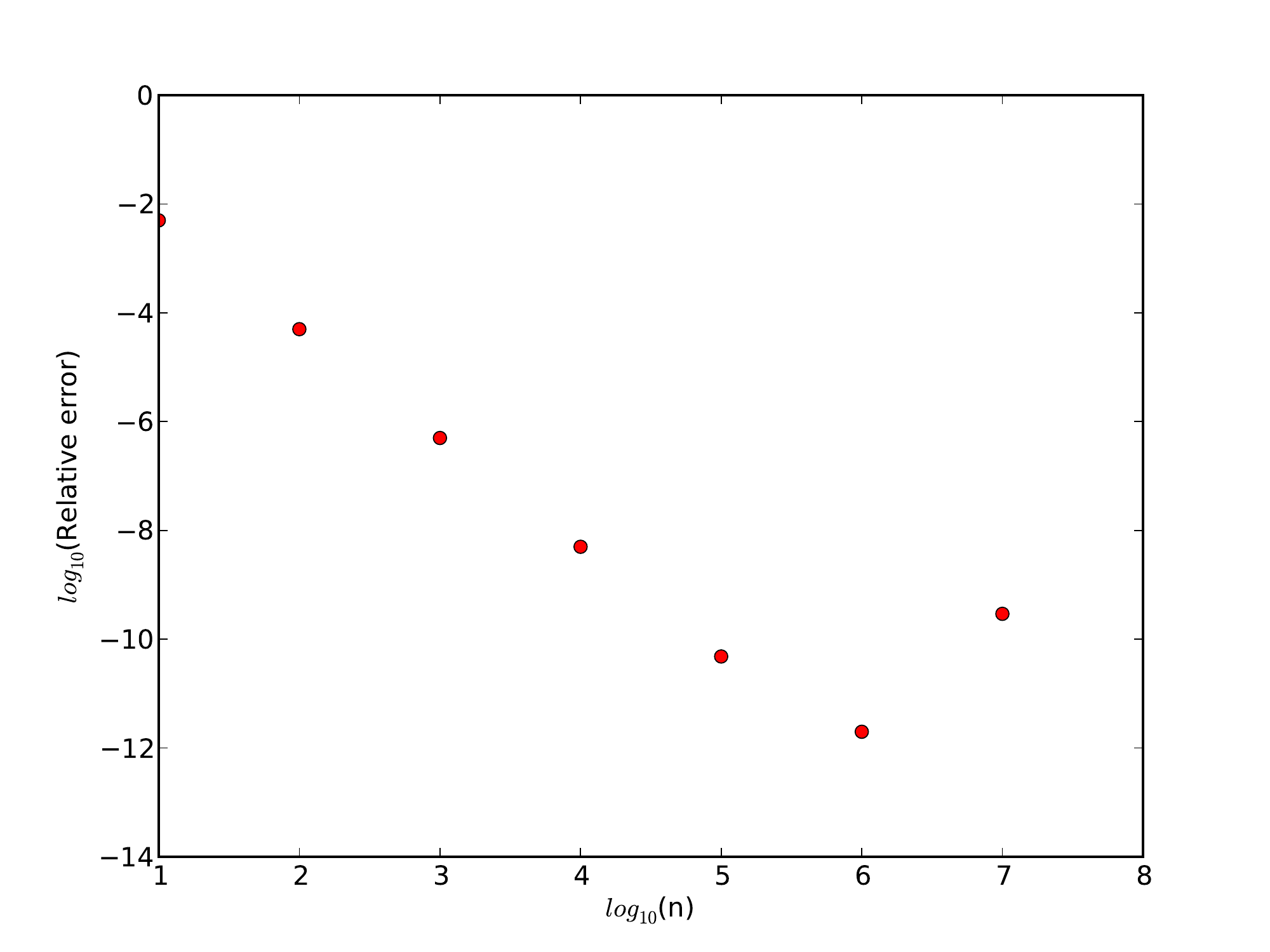}
\caption{Log-log plot of the relative error as function of the number of integration points. Till approximately $n=10^6$, the relative error follows the predicted mathematical error of the trapezoidal rule. For higher numbers of integration points, numerical round off errors and loss of numerical precision give an increasing relative error.}\label{fig:error}
\end{figure}

Coming back to the above learning outcomes, we would like to emphasize that
Python offers an  extremely versatile programming  environment, allowing for
the inclusion of analytical studies in a numerical program. Here we show an
example code with the trapezoidal rule using Python's symbolic package \textbf{SymPy} \cite{SymPy} to evaluate an integral and compute the absolute error
with respect to the numerically evaluated one of the integral
$\int_0^1 dx x^2 = 1/3$. This is in shown in the following code part
\begin{lstlisting}
# define x as a symbol to be used by sympy
x = Symbol('x')
exact = integrate(function(x), (x, 0.0, 1.0))
print("Sympy integration=", exact)
\end{lstlisting}
where we have defined the function to integrate in
the complete Python program that follows here.
\begin{lstlisting}
from math import *
from sympy import *
def Trapez(a,b,f,n):
   h = (b-a)/float(n)
   s = 0
   x = a
   for i in range(1,n,1):
       x = x+h
       s = s+ f(x)
   s = 0.5*(f(a)+f(b)) +s
   return h*s

#  function to compute pi
def function(x):
    return x*x

a = 0.0;  b = 1.0; n = 100
result = Trapez(a,b,function,n)
print("Trapezoidal rule=", result)
# define x as a symbol to be used by sympy
x = Symbol('x')
exact = integrate(function(x), (x, 0.0, 1.0))
print("Sympy integration=", exact)
# Find relative error
print("Relative error", abs((exact-result)/exact))
\end{lstlisting}
The following extended version of the trapezoidal rule allows us to
plot the relative error by comparing with the exact result. By
increasing to $10^8$ points we arrive at a region where numerical
errors start to accumulate, as seen in the figure \ref{fig:error}.
\begin{lstlisting}
from math import log10
import numpy as np
from sympy import Symbol, integrate
import matplotlib.pyplot as plt
# function for the trapezoidal rule
def Trapez(a,b,f,n):
   h = (b-a)/float(n)
   s = 0
   x = a
   for i in range(1,n,1):
       x = x+h
       s = s+ f(x)
   s = 0.5*(f(a)+f(b)) +s
   return h*s
#  function to compute pi
def function(x):
    return x*x
# define integration limits
a = 0.0;  b = 1.0;
# find result from sympy
# define x as a symbol to be used by sympy
x = Symbol('x')
exact = integrate(function(x), (x, a, b))
# set up the arrays for plotting the relative error
n = np.zeros(9); y = np.zeros(9);
# find the relative error as function of integration points
for i in range(1, 8, 1):
    npts = 10**i
    result = Trapez(a,b,function,npts)
    RelativeError = abs((exact-result)/exact)
    n[i] = log10(npts); y[i] = log10(RelativeError);
plt.plot(n,y, 'ro')
plt.xlabel('n')
plt.ylabel('Relative error')
plt.show()
\end{lstlisting}

The last example shows the potential of combining numerical algorithms
with symbolic calculations, allowing thereby students  to
validate their algorithms. With concepts like unit testing, one has
the possibility to test and verify several or all parts of the
code. Validation and verification are then included \emph{naturally}.

The above example allows the student to also test the mathematical
error of the algorithm for the trapezoidal rule by changing the number
of integration points. The students get trained from day one to think
error analysis. Figure \ref{fig:error}  shows clearly the region where the
relative error starts increasing.  The mathematical error which
follows the trapezoidal rule goes as $O(h^2)$ where $h$ is the chosen
numerical step size. It is proportional to the inverse of the number of integration points $n$, that is $h\propto 1/n$.

Before numerical round-off errors and loss of
numerical precision kick in (near $h\sim 10^{-6}$) we see that the
relative error in the log-log plot has a slope which follows the
mathematical error.

There are several additional benefits here. The general learning outcomes on computing can be included as in for example the following ways. We
can easily bake in how to structure a code in terms of
functions and modules, or how to read input data flexibly from the
command line or how to write unit tests etc.  The conventions and
techniques outlined here will save students a lot of time when one
extends incrementally software over time, from simpler to more
complicated problems. In the next subsection we show how algorithms
for solving sets of ordinary differential equations and finding
eigenvalues can be reused in different courses with minor
modifications only.

\subsection{From Mathematics to Physics}

We assume that our students know how to solve and study systems of
ordinary differential with initial conditions only. Later in this
section we will venture into two-point boundary value problems that
can be studied and solved with eigenvalue solvers.

Let us start with initial value problems and ordinary differential
equations. Such equations appear in a wealth of physics
applications. Typical examples students will encounter are the
classical pendulum in a mechanics course, an RLC circuit in the course on
electromagnetism, the modeling of the Solar system in an Astrophysics
course and many other cases.  The essential message is that, with
properly scaled equations, students can use essentially the
same algorithms to solve these problems, either starting with
a simple modified Euler algorithm or a Runge-Kutta class of
algorithms or the so-called Verlet class of algorithms, to mention a few.

The idea is that algorithms students develop and use in one course can be
reused in other courses.  This allows  students to make the
relevant abstractions discussed above, opening up for a much wider
range of applicabilities.

Here we look at two familiar cases from
mechanics and electromagnetism, the equations for the classical
pendulum and those for an RLC circuit.  When properly scaled, these
equations are essentially the same. To scale equations,
either in terms of dimensionless variables or appropriate variables,
is an important aspect which allows the students to see the potential
for abstractions and hopefully see how the problems studied in say a
mechanics course can be transferred to other fields.

The classical pendulum with damping and external force as it could
appear in a mechanics course is given by the following equation of
motion for the angle $\theta$ as function of time $t$
\[
  ml\frac{d^2\theta}{dt^2}+\nu\frac{d\theta}{dt}  +mgsin(\theta)=Acos(\omega t),
\]
where $m$ is its mass, $l$ the length, $\nu$ a damping factor and $A$
the amplitude of an applied external source with frequency
$\omega$. The solution of this type of equations (second-order
differential equations with given initial conditions) is something the
students encounter the first semester thorugh the courses IN1900 and
MAT-INF1100 at the University of Oslo. At Michigan State University
there is now a compulsory course for physics majors that includes many
of these elements.  With this background, students are already familiar with
the numerical solution and visualization of such equations.
If we now
move to a course on electromagnetism, we encounter almost the same
equation for an RLC circuit, namely
\[
L\frac{d^2Q}{dt^2}+\frac{Q}{C}+R\frac{dQ}{dt}=Acos(\omega t),
\]
where $L$ is the inductance, $R$ the applied resistance, $Q$ the
time-dependent charge and $C$ the capacitance.

Let us consider first the classical pendulum equations with damping and an
external force and define the scaled velocity $\hat{v}$ as
\[
   \frac{d\theta}{d\hat{t}} =\hat{v},
\]
where we have defined a dimensionless time variable $\hat{t}$. With
the equation for the velocity we can rewrite the second-order
differential in terms of two coupled first-order differential
equations where the second equation represents the acceleration
\[
   \frac{d\hat{v}}{d\hat{t}} =Acos(\hat{\omega} \hat{t})-\hat{v}\xi-\sin(\theta).
\]
We have scaled the  equations with $\omega_0=\sqrt{g/l}$,
$\hat{t}=\omega_0 t$ and $\xi = mg/\omega_0\nu$. The frequency
$\omega_0$ defines a so-called natural frequency defined by the
gravitational acceleration $g$ and the length of the pendulum $l$. The
frequency $\hat{\omega}= \omega/\omega_0$.  In a similar way, our RLC
circuit can now be rewritten in terms of two coupled first-order
differential equations,
\[
   \frac{dQ}{d\hat{t}} =\hat{I},
\]
and
\[
   \frac{d\hat{I}}{d\hat{t}} =Acos(\hat{\omega} \hat{t})-\hat{I}\xi-Q,
\]
with $\omega_0=1/\sqrt{LC}$, $\hat{t}=\omega_0 t$ and $\xi =
CR\omega_0$. Here we see that the natural frequency is defined in
terms of the physical parameters $L$ and $C$.

The equations are essentially the same, the main differences reside in
the different scaling constants and the introduction of a non-linear
term for the angle $\theta$ in the pendulum equation. The differential
solver the students end up writing in the mechanics course (which
comes normally before the course on electromagnetism) can then be
reused in the electromagnetism course, with a great potential for
further abstraction.

Let us now move to another frequently encountered problem in several
physics courses, namely that of a two-point boundary value problem. In
the examples below we will see again that if the equations are
properly scaled, we can reuse the same algorithm for solving different
physics problems. Here we will start with the equations for a buckling
beam (a case which can be found in a mechanics course or a course on
mathematical methods in physics). Thereafter, with a simple change of
variables and constants, the same problem can be used to study a
quantum mechanical particle confined to move in an infinite potential
well.  By simply changing the diagonal matrix elements of the
discretized differential equation problem, we can study particles that
move in a harmonic oscillator potential or other types of
quantum-mechanical one-body or selected two-body problems.  With
slight modifications to the matrix that results from the
discretization of a second derivative, we can study Poisson's equation
in one dimension, a problem of relevance in
electromagnetism.

Let us start with the buckling beam. This is a two-point boundary
value problem
\[
R \frac{d^2 u(x)}{dx^2} = -F u(x),
\]
where $u(x)$ is the vertical displacement, $R$ is a material specific
constant, $F$ is the applied force and $x \in [0,L]$ with $u(0)=u(L)=0$.
We scale the equation with $x = \rho L$ and $\rho \in [0,1]$ and get
(note that we change from $u(x)$ to $v(\rho)$)
\[
\frac{d^2 v(\rho)}{dx^2} +K v(\rho)=0,
\]
which is, when discretized (see below), nothing but a standard eigenvalue problem with $K=
FL^2/R$. Here we can assume that either the force $F$ or the material
specific rigidity $R$ are unknown.  If we replace $R=-\hbar^2/2m$ and
$-F=\lambda$, we have the quantum mechanical variant for a particle
moving in a well with infinite walls at the endpoints.  The way to
solve these equations numerically is to discretize the second
derivative and the right hand side as
\[
    -\frac{v_{i+1} -2v_i +v_{i-i}}{h^2}=\lambda v_i,
\]
with $i=1,2,\dots, n$. Here $h$ is the step size which is defined by
the number of integration (or mesh) points.  We need to add to this
system the two boundary conditions $v(0) =v_0$ and $v(1) = v_{n+1}$,
although they are not needed in the solution of the equations since
their values are known.  For all integration points $i=1,2,\dots, n$
the set of equations to solve result in a so-called tridiagonal
Toeplitz matrix ( a special case from the discretized second
derivative)
\[
    \mathbf{A} = \frac{1}{h^2}\begin{bmatrix}
                          2 & -1 &  &   &  & \\
                          -1 & 2 & -1 & & & \\
                           & -1 & 2 & -1 & &  \\
                           & \dots   & \dots &\dots   &\dots & \dots \\
                           &   &  &-1  &2& -1 \\
                           &    &  &   &-1 & 2 \\
                      \end{bmatrix}
\]
and with the corresponding vectors $\mathbf{v} = (v_1, v_2,
\dots,v_n)^T$ allows us to rewrite the differential equation as a
standard eigenvalue problem
\[
   \mathbf{A}\mathbf{v} = \lambda\mathbf{v}.
\]
The tridiagonal Toeplitz matrix has analytical eigenpairs,
providing us thereby with an invaluable check on the equations to be
solved.

If we stay with quantum mechanical one-body problems (or
special interacting two-body problems) adding a potential along the
diagonal elements allows us to reuse this problem for many types of physics
cases.  To see this, let us assume we are interested in the solution
of the radial part of Schr\"odinger's equation for one electron. This
equation reads
\[
  -\frac{\hbar^2}{2 m} \left ( \frac{1}{r^2} \frac{d}{dr} r^2
  \frac{d}{dr} - \frac{l (l + 1)}{r^2} \right )R(r)
     + V(r) R(r) = E R(r).
\]
Suppose in our case $V(r)$ is the harmonic oscillator potential
$(1/2)kr^2$ with $k=m\omega^2$ and $E$ is the energy of the harmonic
oscillator in three dimensions.  The oscillator frequency is $\omega$
and the energies are
\[
E_{nl}=  \hbar \omega \left(2n+l+\frac{3}{2}\right),
\]
with $n=0,1,2,\dots$ and $l=0,1,2,\dots$.

Since we have made a transformation to spherical coordinates it means
that $r\in [0,\infty)$. The quantum number $l$ is the orbital momentum
  of the electron.  In order to find analytical solutions for this
  problem, we would substitute $R(r) = (1/r) u(r)$ (which gives
  $u(0)=u(\infty)=0$ and thereby easier boundary conditions) and
  obtain
\[
  -\frac{\hbar^2}{2 m} \frac{d^2}{dr^2} u(r)
       + \left ( V(r) + \frac{l (l + 1)}{r^2}\frac{\hbar^2}{2 m}
                                    \right ) u(r)  = E u(r) .
\]
The boundary conditions are $u(0)=0$ and $u(\infty)=0$.

In order to scale the equations, we introduce a dimensionless variable $\rho = (1/\alpha) r$
where $\alpha$ is a constant with dimension length and get
\[
  -\frac{\hbar^2}{2 m \alpha^2} \frac{d^2}{d\rho^2} v(\rho)
       + \left ( V(\rho) + \frac{l (l + 1)}{\rho^2}
         \frac{\hbar^2}{2 m\alpha^2} \right ) v(\rho)  = E v(\rho) .
\]
Let us choose $l=0$ for the mere sake of simplicity.
Inserting $V(\rho) = (1/2) k \alpha^2\rho^2$ we end up with
\[
  -\frac{\hbar^2}{2 m \alpha^2} \frac{d^2}{d\rho^2} v(\rho)
       + \frac{k}{2} \alpha^2\rho^2v(\rho)  = E v(\rho).
\]
We multiply thereafter with $2m\alpha^2/\hbar^2$ on both sides and obtain
\[
  -\frac{d^2}{d\rho^2} v(\rho)
       + \frac{mk}{\hbar^2} \alpha^4\rho^2v(\rho)  = \frac{2m\alpha^2}{\hbar^2}E v(\rho) .
\]

A natural length scale comes out automatically when scaling. To see this, since $\alpha$ is constant we are left to determine, 
we determine $\alpha$ by requiring that
\[
\frac{mk}{\hbar^2} \alpha^4 = 1.
\]
This defines a natural length scale in terms of the various physical
constants that determine the equation.  The final expression, inserting $k=m\omega^2$ is
\[
\alpha = \left(\frac{\hbar}{m\omega}\right)^{1/2}.
\]
If we were to replace the harmonic oscillator potential with the
attractive Coulomb interaction from the hydrogen atom, the  parameter $\alpha$ would equal the Bohr
radius $a_0$.  This way students see the general properties of a
two-point boundary value problem and can reuse the code they developed
for a mechanics course to the subsequent quantum mechanical course.

Defining
\[
\lambda = \frac{2m\alpha^2}{\hbar^2}E,
\]
we can rewrite Schroedinger's equation as
\[
  -\frac{d^2}{d\rho^2} v(\rho) + \rho^2v(\rho)  = \lambda v(\rho) .
\]
This is similar to the equation for a buckling beam, except for the
potential term.  In three dimensions with our scaling, the eigenvalues for $l=0$ are
$\lambda_0=3,\lambda_1=7,\lambda_2=11,\dots .$

If we  define first the diagonal matrix element
\[
   d_i=\frac{2}{h^2}+V_i,
\]
and the non-diagonal matrix element
\[
   e_i=-\frac{1}{h^2},
\]
we can rewrite the Schr\"oedinger equation as
\[
d_iu_i+e_{i-1}v_{i-1}+e_{i+1}v_{i+1}  = \lambda v_i,
\]
where $v_i$ is unknown and $i=1,2,\dots, n$. We can reformulate the
latter equation as a matrix eigenvalue problem
\[
    \begin{bmatrix} d_1 & e_1 & 0   & 0    & \dots  &0     & 0 \\
                                e_1 & d_2 & e_2 & 0    & \dots  &0     &0 \\
                                0   & e_2 & d_3 & e_3  &0       &\dots & 0\\
                                \dots  & \dots & \dots & \dots  &\dots      &\dots & \dots\\
                                0   & \dots & \dots & \dots  &\dots       &d_{n-1} & e_{n-1}\\
                                0   & \dots & \dots & \dots  &\dots       &e_{n-1} & d_{n}
             \end{bmatrix}      \begin{bmatrix} v_{1} \\
                                                              v_{2} \\
                                                              \dots\\ \dots\\ \dots\\
                                                              v_{n}
             \end{bmatrix}=\lambda \begin{bmatrix}{c} v_{1} \\
                                                              v_{2} \\
                                                              \dots\\ \dots\\ \dots\\
                                                              v_{n}
             \end{bmatrix}
\]
or if we wish to be more detailed, we can write the tridiagonal matrix as
\[
    \begin{bmatrix} \frac{2}{h^2}+V_1 & -\frac{1}{h^2} & 0   & 0    & \dots  &0     & 0 \\
                                -\frac{1}{h^2} & \frac{2}{h^2}+V_2 & -\frac{1}{h^2} & 0    & \dots  &0     &0 \\
                                0   & -\frac{1}{h^2} & \frac{2}{h^2}+V_3 & -\frac{1}{h^2}  &0       &\dots & 0\\
                                \dots  & \dots & \dots & \dots  &\dots      &\dots & \dots\\
                                0   & \dots & \dots & \dots  &\dots       &\frac{2}{h^2}+V_{n-1} & -\frac{1}{h^2}\\
                                0   & \dots & \dots & \dots  &\dots       &-\frac{1}{h^2} & \frac{2}{h^2}+V_{n}
             \end{bmatrix}.
\]

The following Python code sets up the matrix to diagonalize by defining
the minimun and maximum values of $r$ with a maximum value of
integration points. It plots the eigenfunctions of the three lowest
eigenstates.
\begin{lstlisting}
#Program which solves the one-particle Schrodinger equation
#for a potential specified in function
#potential().

from  matplotlib import pyplot as plt
import numpy as np
#Function for initialization of parameters
def initialize():
    RMin = 0.0
    RMax = 10.0
    lOrbital = 0
    Dim = 400
    return RMin, RMax, lOrbital, Dim
# Harmonic oscillator potential
def potential(r):
    return 0.5*r*r

#Get the boundary, orbital momentum and number of integration points
RMin, RMax, lOrbital, Dim = initialize()

#Initialize constants
Step    = RMax/(Dim)
DiagConst = 2.0/ (Step*Step)
NondiagConst =  -1.0 / (Step*Step)
OrbitalFactor = lOrbital * (lOrbital + 1.0)

#Calculate array of potential values
v = np.zeros(Dim)
r = np.linspace(RMin,RMax,Dim)
for i in range(Dim):
    r[i] = RMin + (i+1) * Step;
    v[i] = potential(r[i]) + OrbitalFactor/(r[i]*r[i]);

#Setting up a tridiagonal matrix and finding eigenvectors and eigenvalues
Matrix = np.zeros((Dim,Dim))
Matrix[0,0] = DiagConst + v[0];
Matrix[0,1] = NondiagConst;
for i in xrange(1,Dim-1):
    Matrix[i,i-1]  = NondiagConst;
    Matrix[i,i]    = DiagConst + v[i];
    Matrix[i,i+1]  = NondiagConst;
Matrix[Dim-1,Dim-2] = NondiagConst;
Matrix[Dim-1,Dim-1] = DiagConst + v[Dim-1];
# diagonalize and obtain eigenvalues, not necessarily sorted
EigValues, EigVectors = np.linalg.eig(Matrix)
# sort eigenvectors and eigenvalues
permute = EigValues.argsort()
EigValues = EigValues[permute]
EigVectors = EigVectors[:,permute]
# now plot the results for the three lowest lying eigenstates
for i in range(3):
    print(EigValues[i])
FirstEigvector = EigVectors[:,0]
SecondEigvector = EigVectors[:,1]
ThirdEigvector = EigVectors[:,2]
plt.plot(r, FirstEigvector**2 ,'b-',r, SecondEigvector**2 ,'g-',r, ThirdEigvector**2 ,'r-')
plt.axis([0,4.6,0.0, 0.025])
plt.xlabel(r'$r$')
plt.ylabel(r'Radial probability $r^2|R(r)|^2$')
plt.title(r'Radial probability distributions for three lowest-lying states')
plt.savefig('eigenvector.pdf')
plt.show()
\end{lstlisting}

\begin{figure}
\includegraphics[scale=0.8]{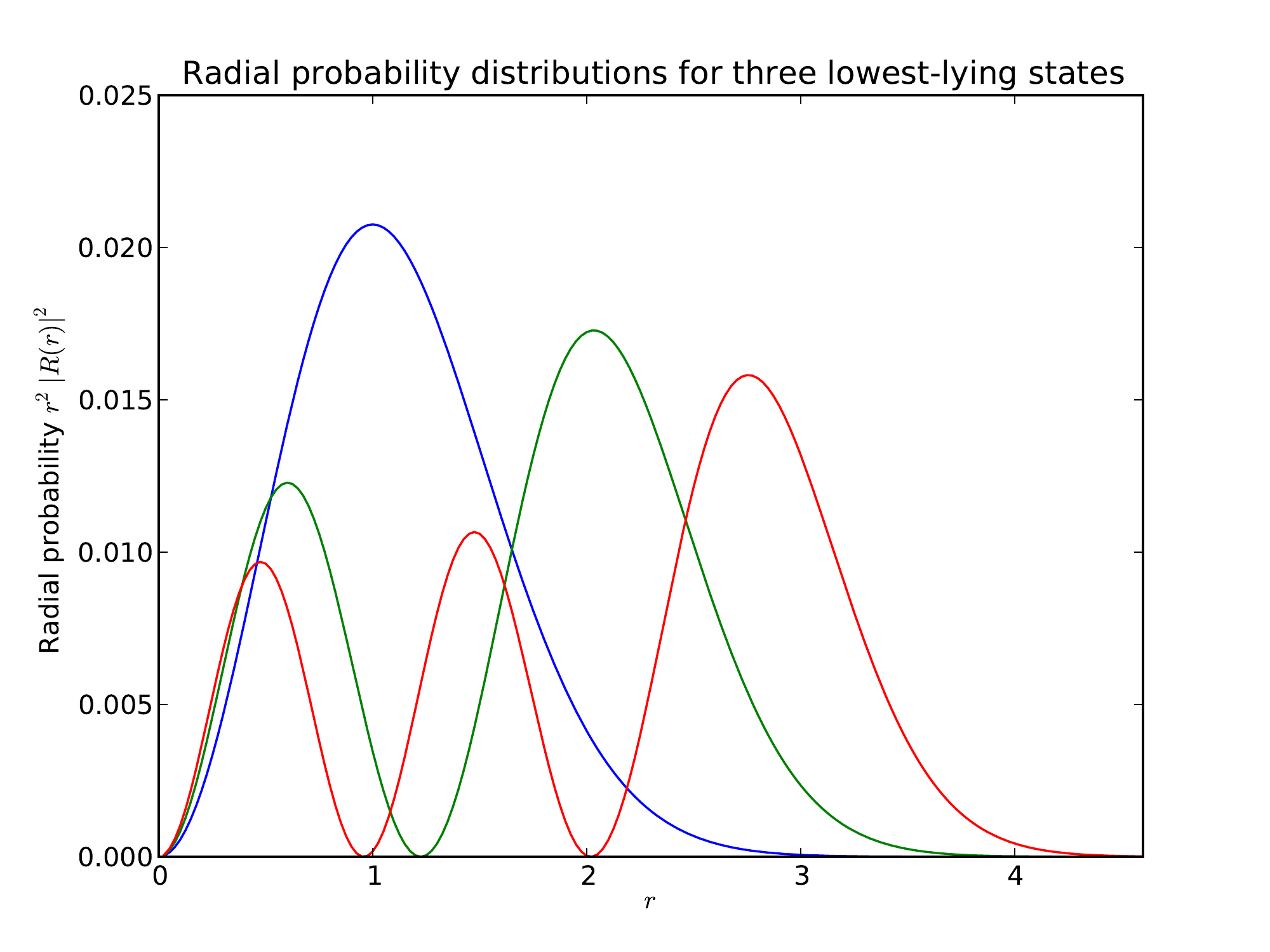}
\caption{Plot of the eigenfunctions of the three lowest-lying eigenvalues for a harmonic oscillator problem in three dimensions. The students can easily change the type of potential and explore the physics that arises from these potentials.}\label{fig:eigenvector}
\end{figure}

The last example shows the potential of combining numerical algorithms with analytical results (or eventually symbolic calculations), allowing thereby students to test their physics understanding. One can easily switch to other potentials by simply redefining the potential function. For example, a finite box potential can easily be defined as
\begin{lstlisting}
# Finite depth and range box potential, with strength V and range a
def potential(r):
    if r >= 0.0 and r <= 10.0:
        V = -0.05
    else:
        V =0.0
    return V
\end{lstlisting}
Thereafter, the students can explore the role of the potential depth
and the range of the potential. Analyzing the eigenvectors gives
additional information about the spatial degrees of freedom in terms
of different potentials.  The possibility to visualize the results immediately, as shown in figure \ref{fig:eigenvector}, aids in providing students with a deeper understanding of the relevant physics.

This example contains also many of the
computing learning outcomes we discussed above, in addition to those
related to the physics of a particular system. We see that, by proper
scaling, the students can make further abstractions and explore other
physics cases easily where no analytical solutions are known. With
unit testing and analytical results they can validate and verify their
algorithms.

The above example allows the student to test the mathematical error of
the algorithm for the eigenvalue solver by simply changing the number
of integration points. Again, as discussed above in connection with
the trapezoidal rule, the students get trained to develop an
understanding of the error analysis and where things can go wrong. The
algorithm can be tailored to any kind of one-particle problem used in
quantum mechanics.

A simple rewrite allows for the reuse in linear algebra problems for
solution of say Poisson's equation in electromagnetism, or the
diffusion equation in one dimension. To see this and how the same matrix can be used in a course in electromagnetism, let us consider
Poisson's equation.
We assume that
the electrostatic potential $\Phi$ is generated by a localized charge
distribution $\rho (\mathbf{r})$.   In three dimensions
the pertinent equation reads
\[
\nabla^2 \Phi = -4\pi \rho (\mathbf{r}).
\]
With a spherically symmetric potential $\Phi$ and charge distribution $\rho (\mathbf{r})$ and using spherical coordinates,  the relevant
equation to solve
simplifies to a one-dimensional equation in $r$, namely
\[
\frac{1}{r^2}\frac{d}{dr}\left(r^2\frac{d\Phi}{dr}\right) = -4\pi \rho(r),
\]
which can be rewritten via a substitution $\Phi(r)= \phi(r)/r$ as
\[
\frac{d^2\phi}{dr^2}= -4\pi r\rho(r).
\]
The inhomogeneous term $f$ or source term is given by the charge distribution
$\rho$  multiplied by $r$ and the constant $-4\pi$.

We can  rewrite this equation by letting $\phi\rightarrow u$ and
$r\rightarrow x$.  Scaling again the equations and replacing the right hand side with a function $f(x)$, we can rewrite the
equation as
\[
-u''(x) = f(x).
\]
Our scaling gives us again $x\in [0,1]$ and the two-point boundary value problem
with $u(0)=u(1)=0$. With $n+1$ integration points and
the step length defined as $h=1/(n)$ and replacing the continuous function $u$ with its discretized version $v$, we get
the following equation
\begin{equation*}
   -\frac{v_{i+1}+v_{i-1}-2v_i}{h^2} = f_i  \hspace{0.5cm} \mathrm{for} \hspace{0.1cm} i=1,\dots, n,
\end{equation*}
where $f_i=f(x_i)$.
Bringing up again the tridiagonal Toeplitz matrix,
\[
    \mathbf{A} = \frac{1}{h^2}\begin{bmatrix}
                           2& -1& 0 &\dots   & \dots &0 \\
                           -1 & 2 & -1 &0 &\dots &\dots \\
                           0&-1 &2 & -1 & 0 & \dots \\
                           & \dots   & \dots &\dots   &\dots & \dots \\
                           0&\dots   &  &-1 &2& -1 \\
                           0&\dots    &  & 0  &-1 & 2 \\
                      \end{bmatrix},
\]
our problem becomes now a classical linear algebra problem
\[
\mathbf{A}\mathbf{v}=\mathbf{f},
\]
with the unknown function $\mathbf{v}$. Using standard LU
decomposition algorithms \cite{GolubVanLoan} (here one can use
the so-called Thomas algorithm which reduces the number of floating
point operations to $O(n)$) one can easily find the solution to this
problem.

These examples demonstrate how one can, with a discretized second
derivative, solve physics problems that arise in different
undergraduate courses using standard linear algebra and eigenvalue
algorithms and ordinary differential equations, allowing thereby
teachers to focus on the interesting physics. Many of these problems
can easily be linked up with ongoing research. This opens up for many
interesting perspectives in physics education. We can bring in at a
much earlier stage in our education basic research elements and
perhaps even link with ongoing research during the first year of
undergraduate studies.

Instead of focusing on tricks and mathematical manipulations to solve
the continuous problems for those few case where an analytical
solution can be found, the discretization of the continuous problem
opens up for studies of many more interesting and realistic problems.
However, we have seen that in order to verify and validate our codes,
the existence of analytical solutions offer us an invaluable test of
our algorithms and programs. The analytical results can either be
included explicitely or via symbolic software like Python's Sympy package.
Thus, computing stands indeed for solving scientific problems using
all possible tools, including symbolic computing, computers and
numerical algorithms, numerical experiments (as well as real
experiments if possible) and analytical paper and pencil solutions.

The cases we have presented here represent only a limited set of
examples. A longer version of this article, with more examples and
details on assessments programs, is under preparation as a textbook
\cite{DannyMortenBook}.  The possible learning outcomes we defined for
various physics courses are often based on the above simple
discretization. With basic knowledge on how to solve linear algebra
problems, eigenvalue porblems and differential equations, topics
normally taught in mathematics and computational science courses, we
can offer our students a much more challenging and interesting
education. Furthermore, we give our students the competencies which are
required by future employers, either in the private or the public
sector.

\section{Conclusions and Perspectives}

In this contribution, we have outlined some of the basic elements that we
feel are necessary to address in order to introduce computing in
various undergraduate physics courses.  Some of the conclusions we
would like to emphasize include a proper  definition of computing,
the development of learning outcomes that apply to both computational science, mathematics,  and
physics courses as well as  proper assessment programs.

Collaboration across departments is necessary in order to achieve a
synchronization between various topics and learning outcomes, as well
as an early introduction to programming. Many universities
require such courses as part of a physics degree. Coordinating such
a programming course with mathematics courses and other science
courses results in a better coordination of both learning outcomes and
computing skills and abilities. The experiences we have drawn from the
University of Oslo and Michigan State University show that an early
and compulsory programming course, which includes central scientific elements, is
important in order to integrate properly a computational perspective
in our physics education.

The benefits are many, in particular it allows us to make our research
more visible in early undergraduate physics courses, enhancing
research-based teaching with the possibility to focus more on
understanding and increased insight.  It gives also our candidates the
skills and abilities that are requested by society at large, both from
the private and the public sectors. With computing, we emphasize a broader and
more up-to-date education with a problem-based orientation, often
requested by potential employers.  Furthermore, our experiences from the both universities
indicate that a discussion of computing across disciplines results in an increased
impetus for broad cooperation in teaching and a broader focus on university pedagogical topics.

We are now in the process of developing computing learning outcomes with
examples for central physics courses. Together with a research based assessment program, we will be able to answer central questions like whether the introduction of computing increases a student's insights and understanding of the underlying physics.

\begin{acknowledgement}
MHJ's work is supported by U.S. National Science Foundation Grant No.~PHY-1404159.
MDC's work is supported by U.S. National Science Foundation Grants Nos.~DRL-1741575, DUE-1725520, DUE-1524128, DUE-1504786, and DUE-1431776.
Both authors acknowledge support from the recently established Center for Computing in Science Education, University of Oslo, Norway.
\end{acknowledgement}


\end{document}